\documentclass[12pt,letterpaper]{article}
\usepackage{jheppub}
\usepackage{hyperref}
\usepackage[T1]{fontenc}
\usepackage[latin9]{inputenc}
\usepackage{color}
\usepackage{amstext}
\usepackage{amssymb}
\usepackage{stmaryrd}
\usepackage{graphicx}
\usepackage{esint}
\usepackage{cancel}
\usepackage{ulem}
\usepackage{afterpage}

\renewcommand{\arraystretch}{1.1}

\makeatletter
\providecommand{\tabularnewline}{\\}
\makeatother

\newcommand{\be}{\begin{equation}}
\newcommand{\ee}{\end{equation}}

\newmuskip\pFqmuskip

\newcommand*\pFq[6][8]{%
	\begingroup 
	\pFqmuskip=#1mu\relax
	\mathcode`\,=\string"8000
	\begingroup\lccode`\~=`\,
	\lowercase{\endgroup\let~}\pFqcomma
	{}_{#2}F_{#3}{\left[\genfrac..{0pt}{}{#4}{#5};#6\right]}%
	\endgroup
}
\newcommand{\pFqcomma}{\mskip\pFqmuskip}

\title{RG stability of integrable fishnet models}

\author[a]{Ohad Mamroud}
\author[a]{ and Gen\'is Torrents}

\affiliation[a]{Department of Particle Physics and Astrophysics,
Weizmann Institute of Science, Rehovot 7610001, Israel}

\emailAdd{ohad.mamroud@weizmann.ac.il}
\emailAdd{genis.torrents@weizmann.ac.il}

\abstract{We address the question of perturbative consistency in the scalar
fishnet models presented by Caetano, G\"urdo\u{g}an and Kazakov\cite{Gurdogan:2015csr, Caetano:2016ydc}.
We argue that their 3-dimensional $\phi^{6}$ fishnet model becomes perturbatively stable under renormalization in the large $N$ limit, in
contrast to what happens in their 4-dimensional $\phi^{4}$ fishnet
model, in which double trace terms are known to be generated by the RG flow. We point
out that there is a direct way to modify this second theory that protects
it from such corrections. Additionally, we observe that the 6-dimensional
$\phi^{3}$ Lagrangian that spans an hexagonal integrable scalar fishnet
is consistent at the perturbative level as well. The nontriviality
and simplicity of this last model is illustrated by computing the anomalous dimensions of its $\text{tr}\phi_i \phi_j$ operators to all perturbative orders.}

\begin{document}
\maketitle

\section{Introduction}

For many decades, the inherent complexity to general nonperturbative QFT analysis has encouraged researchers to focus on two exceptionally convenient toy models: $\mathcal{N}=4$ SYM and ABJM. The study of these theories in the strong coupling regime makes use of two 
of their main virtues - their expected
holographic realization \cite{Aharony:1999ti} and their integrability in
the 't Hooft limit \cite{Beisert:2010jr}. Despite the lack of a formal proof for these properties, the amount of evidence supporting
them both perturbatively and at the nonperturbative level is overwhelming.
The abundance of nontrivial results that stems from the application of holographic and integrability techniques to these theories has located them,
and especially $\mathcal{N}=4$ SYM, in a privileged position as prospective
departure points in the exploration of the nonperturbative landscape.

Recently, several efforts have been devoted to the characterization
of the integrability-preserving continuous deformations of these maximally
supersymmetric models\cite{Lunin:2005jy, Frolov:2005dj, Beisert:2005if, Chen:2016geo, Imeroni:2008cr},
usually called $\gamma$-deformations in the literature. These type
of deformations can be implemented in the holographic setting using
a sequence of up to three independent T-duality, shift, T-duality
(TsT) transformations. Equivalently, they are obtained in the field theoretic side by replacing the commutators of the theory with a Moyal $\star$-product
with three twist parameters $\gamma_{i}$ directly related to the
shifts in the holographic picture. 

In this context an unexpected observation was made \cite{Gurdogan:2015csr, Caetano:2016ydc}:
A particular limit of $\gamma$-deformation that was dubbed the ``double
scaling limit'' in \cite{Gurdogan:2015csr} gives rise to scalar matrix
models with classically marginal couplings and a surprisingly simple
diagrammatic structure. More precisely, in the large $N$ limit of
these models the only nontrivial dependence on the coupling for most
quantities arises at the perturbative level from fishnet Feynman diagrams,
i.e., arrangements of propagators in the shape of a regular lattice, which is made of triangular and square cells in the cases of ABJM and $
\mathcal{N}=4$ SYM respectively. Both types of fishnet have been explicitly
proven to be amenable to integrability \cite{Zamolodchikov:1980mb},
although in a framework that is not trivially connected to the inherited
integrability from $\mathcal{N}=4$ SYM. The understanding on whether (or how) the two types of integrability are related could shed light on the field-theoretic origin of integrability in $\mathcal{N}=4$ SYM. In addition to this interest, the simplicity of the models raises hope that finite and strong coupling calculations could be carried out for them.

The aforementioned construction has, however, an important weakness:
Generic $\gamma$-deformations \cite{Fokken:2013aea, Fokken:2014soa},
and in particular the fishnet model for the $\mathcal{N}=4$ $\gamma$-deformation
\cite{Sieg:2016vap}, are unstable under RG flow, and therefore need
to be corrected by adding double trace operators to the Lagrangian.
Such terms are not unprecedented in deformations of $\mathcal{N}=4$
or ABJM: similar double trace corrections arise in the discrete deformations
implemented by orbifolding the compact space in the holographic dual
\cite{Kachru:1998ys, Lawrence:1998ja, Bershadsky:1998mb, Bershadsky:1998cb}. In the
context of $AdS_5$ orbifolds it has been argued that their $\beta$ functions generically do not vanish on the real axis \cite{Dymarsky:2005nc, Dymarsky:2005uh},
precluding the existence of a perturbative stabilization of the RG
flow in any theory where neither these double trace corrections
nor their flow are protected by additional symmetries\footnote{To the knowledge of the authors, only two examples are known to be
free from such corrections: the supersymmetric $\gamma$ deformation,
known as $\beta$-deformation \cite{Lunin:2005jy}, which implements
a particular Leigh-Strassler deformation of $\mathcal{N}=4$ where
the double trace operators do not flow, and the orbifold of type
0B suggested by Pedro Liendo, which is free from this type of operators
by construction \cite{Liendo:2011da}.}. 

It remains unclear up to which point the RG flow stability condition
can restrict analytically continued $\gamma$-deformations (for which
unitarity has been relaxed): one might expect in these cases the possibility
to fine tune the complex coupling to double trace operators at any
perturbative order \cite{Sieg:2016vap}. Be that as it may, in a stable
theory defined by a double scaling limit the presence of such double
trace elements at planar level significantly convulses the full diagrammatic
of the problem, and restricts the applicability of integrability results
to a narrow set of observables. Allegedly, one can still obtain useful
results \cite{Gurdogan:2015csr} for this protected sector. Nevertheless, many of the
prospective applications of the fishnet models, whose interest may reach beyond the scope of $\gamma$ deformations due to their perturbative simplicity and nontriviality, disappear as soon as additional couplings are turned on in the Lagrangian.

In this note we present three results that show how this difficulty can be circumvented. Our first contribution in this regard is to prove that the fishnet
model that stems out of ABJM is not afflicted with such perturbative
counterterms in the large $N$ limit. In addition, we introduce a modification
in the corresponding limit of $\mathcal{N}=4$ that can preclude this class of terms as well. Separately, we consider the set of hexagonal
fishnet diagrams, which are the only class of integrable fishnet diagrams
presented in \cite{Zamolodchikov:1980mb} for which a fishnet theory
had not been implemented yet. We present a 6-dimensional Lagrangian that produces this hexagonal theory, and prove its perturbative renormalizability at the planar limit.

Most of these results follow directly from the analysis of section
\ref{sec:Divergs}, where each fishnet model with a regular lattice is considered separately in order to explicitly identify all its possible RG flow instabilities. For the sake of generality, our analysis has not been restricted to scalar models; we allow for fermionic lines as well. Among the considered models, the triangular and hexagonal models with scalar fields are singled out as the only examples where double trace corrections are not turned on by loop corrections. In fact, the hexagonal model would also radiate $\text{tr}^{2}\phi$ counterterms, but when this model is realized
as a double scaling limit of a gauge theory these double trace terms
are prevented by Gauss's law. 

Section \ref{sec:Double traces} focuses on double trace operators.
After briefly discussing which observables would be sensitive to their
presence at the planar limit provided a hypothetical RG stabilization
for them was found, we switch to a suggestive observation: Simple
deformations of the scalar square fishnet model can perturbatively elude
radiating double trace terms by means of a super-selection process that we call ``refinement''. It is implemented by equipping  each type of field  with an additional flavour index
and making the insertions of interacting vertices
shift this index cyclically. Despite reducing the amount of nontrivial
observables and the set of orders at which they receive corrections,
all the diagrams that contribute to the amplitudes of interest of the refined theory will
still be governed by square fishnets.

The paper concludes with section \ref{sec:AllOrders} which illustrates
the strong implications of the diagrammatic simplicity arising in
these models: Without even relying on integrability techniques, it is possible to re-sum the perturbative expansion for the anomalous dimension of $\mbox{tr}\phi_i\phi_j$ to all orders. The re-summation
matches the predictions at finite order, which are discussed in the appendix to the paper. The results are potentially useful from the integrability viewpoint. As we explain in section \ref{subsec:Integrability}, our re-summation procedure has a natural interpretation from the perspective of \cite{Zamolodchikov:1980mb}, but computes quantities that involve nontrivial finite size corrections in the planar limit \cite{Sieg:2005kd}.

\begin{figure}
	\begin{centering}
		\includegraphics[width=1\textwidth]{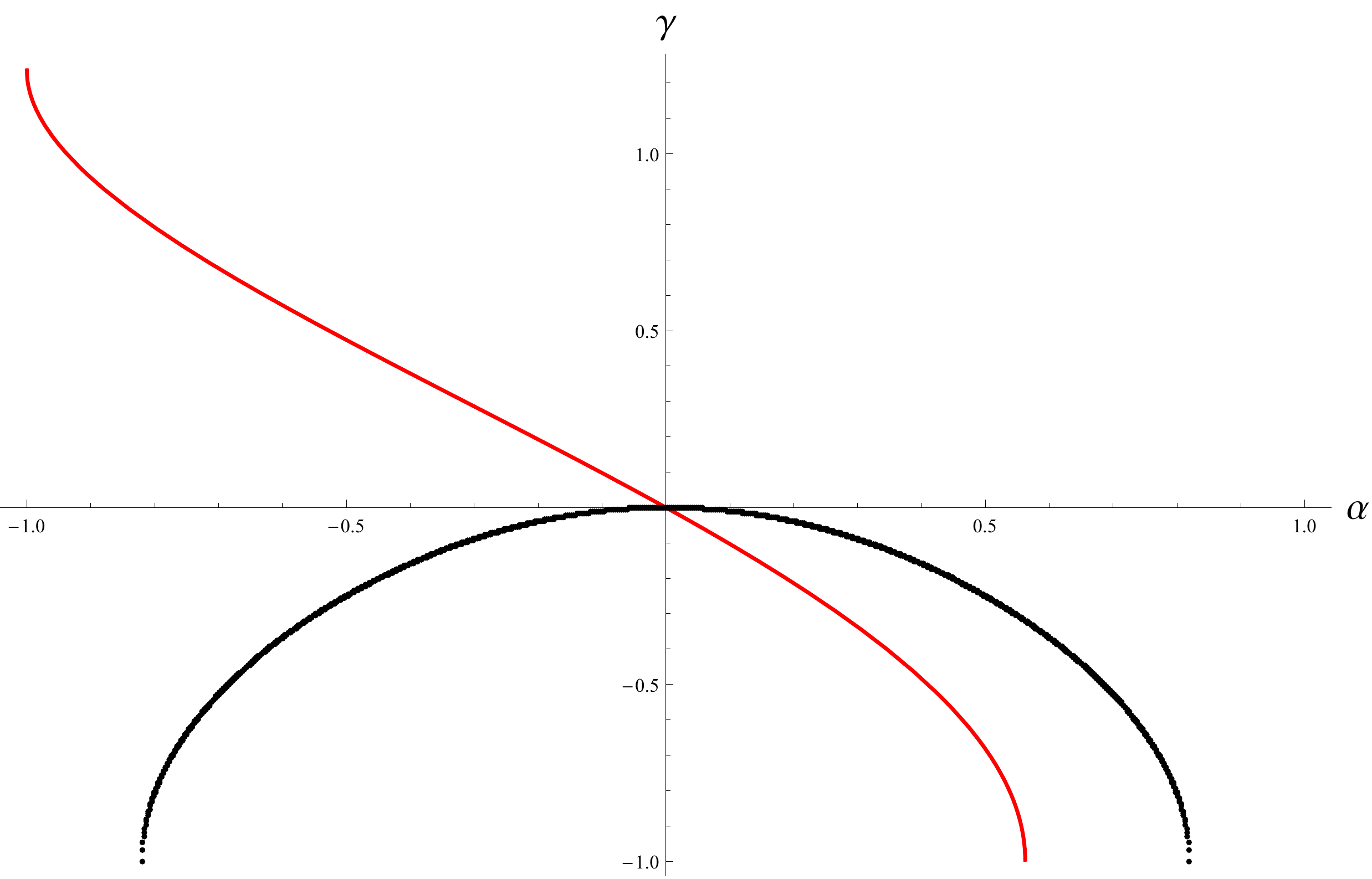}
		\par\end{centering}
	\caption{\label{fig:Summary}Anomalous dimension of $\mbox{tr}\phi_i^2$ (black) and $\mbox{tr}\phi_i\phi_{j\neq i}$ (red) in the hexagonal fishnet model as a function of the only meaningful combination of the couplings, $\alpha$.}
\end{figure}

Figure \ref{fig:Summary} displays the branches of solutions for $\mbox{tr}\phi_i^2$ and $\mbox{tr}\phi_i\phi_{j\neq i}$ that are continuously connected to their classical values.
Interestingly, there is a parametric window,
\begin{equation}
\alpha\in\left(-\frac{3}{4}\sqrt{\frac{3}{19-18C}},\frac{9}{16}\right)
\end{equation}
where $C$ denotes the Catalan constant, in which both real branches simultaneously exist. The endpoints of these branches correspond to points where two real branches of solutions collide with each other. Speculatively, they could signal level crossings, instabilities or phase transitions for the system, despite there is a priori no reason to discard the possibility that these points lie already beyond the validity regime of our analysis because of other phase transitions or instabilities. In fact, it is tantalizing to attribute the three endpoints with $\gamma=-1$ to the fact that the square of the corresponding operator (which is of double trace type and irrelevant at $\alpha=0$) would become marginal with this anomalous dimension. Conversely, the presence of complex solutions for $\gamma$ beyond those points does not necessarily lead to inconsistencies. To the knowledge of the authors no principle is known which protects the anomalous dimensions from becoming complex in nonunitary theories.

In view of the RG stability of the examples considered along the paper, and of the finiteness and reality of the anomalous dimension computed in its last section, it is plausible to expect that some of the models under scrutiny have a well-defined nonperturbative completion. Since their actions are complex, it is natural to attempt such completion by taking the defining path integral of the theory as an integration over cycles given by Lefschetz thimbles. This framework has been shown in the literature to provide other complex theories with well-defined meanings: it links the stokes behaviour in the complexification of couplings to the accumulation of Yang-Lee zeroes in transition phenomena \cite{Guralnik:2007rx, Kanazawa:2014qma}, and potentially solves the ambiguity and ill-definiteness on theories with potentials unbounded from below \cite{Witten:2010cx}. In fact, this framework would constitute the appropriate setting to clarify the meaning of the different branches of solutions we find for $\gamma$, and understand what happens when they meet and whether it signifies a mixing with another operator. The authors plan to work on this question in the near future.


There are several other ways in which the analysis of the present paper can be extended: Two natural questions are whether the integrability presented in \cite{Zamolodchikov:1980mb} can be applied to other cases with fishnet structure, and whether the diagrammatic simplicity of the models considered allows for any simplification beyond planar level. In addition, it would be very interesting to clarify whether the proposed hexagonal and refined square fishnet models can be implemented
as an integrability-preserving double scaling limit of a supersymmetric construction, and, if that is the case, how is the presence of double
trace corrections avoided in them and what is their holographic description.

\section{\label{sec:Divergs}Radiated loop corrections in regular fishnet models}

In this section we generalize the set of Lagrangians presented by Caetano, G\"urdo\u{g}an and Kazakov \cite{Gurdogan:2015csr, Caetano:2016ydc} and discuss under which circumstances the resulting construction is stable under the RG flow, and therefore protected from the appearance of double trace operators.

\begin{figure}
	\begin{centering}
		\includegraphics[width=1\textwidth]{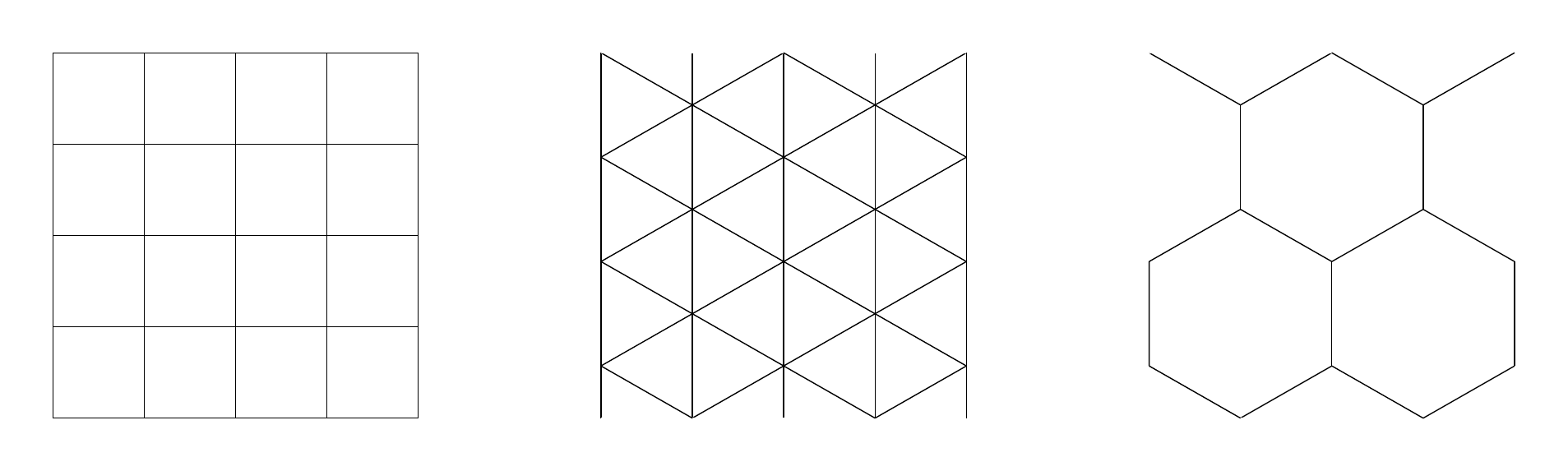}
		\par\end{centering}
	\caption{\label{fig:Lattices}The three types of fishnets that have been proven
		integrable for, respectively, $4$, $3$ and $6$ dimensional
		target spaces happen to coincide with the three regular tilings of
		flat 2-dimensional space.}
\end{figure}

The fishnet models are named that way because in the computation of the anomalous dimension of their single trace operators only a particular set of planar Feynman diagrams contribute: those that look like periodic planar lattices. For the sake of definiteness, in what follows we will restrict ourselves to the cases where these lattices are regular tessellations of the plane, of the types shown in figure \ref{fig:Lattices}. 
In fact, this particular case of diagrams are known to be amenable to integrable methods when all their lines are scalar \cite{Zamolodchikov:1980mb}, but even when that is not the case the simplicity of the perturbative expansion of the theory makes it a convenient toy model for perturbative problems and, plausibly, for resurgent methods.

If we do not take into account the need for renormalizability it is straightforward to engineer Lagrangians that produce only the desired planar fishnets. For the simple examples of figure \ref{fig:Lattices}, we proceed as follows: First, we associate a different species of field to every possible direction on the diagram. We choose these fields to be complex scalars or Dirac fermions in order to have oriented lines that distinguish in which vertex they begin and in which they end. In any case, we will suppose that these fields transform in the adjoint representation of some Lie group, and that this group has a well-defined large $N$ limit in which the algebra of fusion and fission of traces becomes that of $U\left(N\right)$, i.e.
\begin{equation}
\left(T^{i}\right)_{b}^{a}\left(T_{i}\right)_{d}^{c}\propto\delta_{d}^{a}\delta_{c}^{b}+\ldots\rightarrow
\begin{array}{c}
\mbox{tr}\left(T^{i} A \right)\left(T_{i}B\right)\propto \mbox{tr}\left(AB\right)+\ldots\\
\mbox{tr}\left(T^{i} A T_{i}B\right)\propto \mbox{tr}\left(A\right)\mbox{tr}\left(B\right)+\ldots
\end{array}
\label{eq:Traces}
\end{equation}
where the symbol ``$\ldots$'' denotes terms that produce subleading contributions of order $\mathcal{O}\left(N^{-1}\right)$. The construction is completed by writing down interacting terms in the Lagrangian that mimic the chiral color ordering of the fishnet vertices with the chosen prescriptions. With this procedure, we obtain the following list\footnote{The hexagonal fishnet theory with a fermionic species has not been included in the list, because it contains massless bosons, and therefore the tadpole diagrams that result from self-contractions are sensitive to the IR regulator.} of fishnet candidates with marginal interacting terms:
\begin{equation}
\left\{ \begin{array}{c}
\mathcal{S}_{+}^{\left(\text{proposed}\right)}=
N\int d^{4}x\text{Tr}\left(\frac{1}{2}\sum_{i=1}^{2}\left|\partial_{\mu}\phi^{i}\right|^{2}+\xi\phi_{1}\phi_{2}\phi_{1}^{\dagger}\phi_{2}^{\dagger}\right)\\
\mathcal{S}_{+,f}^{\left(\text{proposed}\right)}=
N\int d^{3}x\text{Tr}\left(\frac{1}{2}\left|\partial_{\mu}\phi\right|^{2}+\bar{\psi}\cancel{\partial}\psi+\xi\bar{\psi}\phi\psi\phi^{\dagger}\right)\\
\mathcal{S}_{+,2f}^{\left(\text{proposed}\right)}=N\int d^{2}x\text{Tr}\left(\sum_{i=1}^{2}\bar{\psi}_{i}\cancel{\partial}\psi_{i}\right)+
N\xi \bar{\psi}_{1} \bar{\psi}_{2} \psi_{1} \psi_{2}\\
\mathcal{S}_{\ast}^{\left(\text{proposed}\right)}=N\int d^{3}x\text{Tr}\left(\frac{1}{2}\sum_{i=1}^{3}\left|\partial_{\mu}\phi^{i}\right|^{2}+\xi\phi_{1}\phi_{2}\phi_{3}\phi_{1}^{\dagger}\phi_{2}^{\dagger}\phi_{3}^{\dagger}\right)\\
\mathcal{S}_{\Yup\Ydown}^{\left(\text{proposed}\right)}=N\int d^{6}x\text{Tr}\left(\frac{1}{2}\sum_{i=1}^{3}\left|\partial_{\mu}\phi^{i}\right|^{2}+\xi_{1}\phi_{1}^{\dagger}\phi_{2}\phi_{3}+\xi_{2}\phi_{1}\phi_{2}^{\dagger}\phi_{3}^{\dagger}\right)\\
\mathcal{S}_{\Yup\Ydown,2f}^{\left(\text{proposed}\right)}=N\int d^{4}x\text{Tr}\left(\frac{1}{2}\left|\partial_{\mu}\phi^{i}\right|^{2}+\sum_{i=1}^{2}\bar{\psi}_{i}\cancel{\partial}\psi_{i}+\xi_{1}\bar{\psi}_{1}\psi_{2}\phi+\xi_{2}\bar{\psi}_{2}\phi\psi\right)
\end{array}\right\} \label{eq:S1ProposedLagrangians}
\end{equation}
We shall remain unspecific in this paper about the spinor index structure. The upcoming discussion in this section is phrased in terms of dimensional analysis and the identification of planar diagrams, and therefore insensitive to this choice.

Most of the theories defined in \eqref{eq:S1ProposedLagrangians}
are not consistent as they stand: the proposed Lagrangians have not
been written from the exhaustive list of relevant and marginal operators
protected by some symmetry, and therefore it is plausible (and, indeed,
the case) that the radiative corrections source other terms in the
action along the RG flow. In order to identify which contribute at
the planar level we need to focus on the set of superficially divergent
planar diagrams and check (when their Feynman integral is divergent)
their color structure. 

Before we perform this task, however, it is convenient to emphasize the reasons underlying the diagrammatic simplicity of these models. Notice first that each interacting Lagrangian in \eqref{eq:S1ProposedLagrangians} contains only single trace terms, and that in their entire sum each field and antifield appears only once. This uniquely fixes the way in which vertices can get contracted at the tree level. Moreover, it gives us access to the full list of connected diagrams, which can always be built from the appropriate tree via the contraction of a subset of its external legs.

This idea is of great help in the analysis of the large $N$ limit of the theory, when it is combined with the conventional $N$-power counting that follows from \eqref{eq:Traces}. When we draw a finite tree diagram on the plane with the proper color order of fields at each vertex, we have a cyclical ordered sequence of external legs along the perimeter of the figure, i.e., a single trace operator. Any conjugated pair in this sequence can be contracted, but, in doing so, as \eqref{eq:Traces} indicates, the single trace is divided in two single trace factors. They correspond to the external legs found on either side of the line that represents the contraction. Further contractions within each factor are still possible at the planar level, but any contraction between different blocks explicitly breaks the planarity of the diagram. As a consequence, when neighbouring contractions are not possible, every contraction will increase the amount of single trace factors in the color structure of the planar diagram. This fact importantly decimates the amount of planar diagrams we can construct with a single or double trace structure, and therefore it is a very convenient tool to discuss both perturbative divergences and single trace anomalous dimensions.

\begin{figure}
	\begin{centering}
		\includegraphics[width=0.8\textwidth]{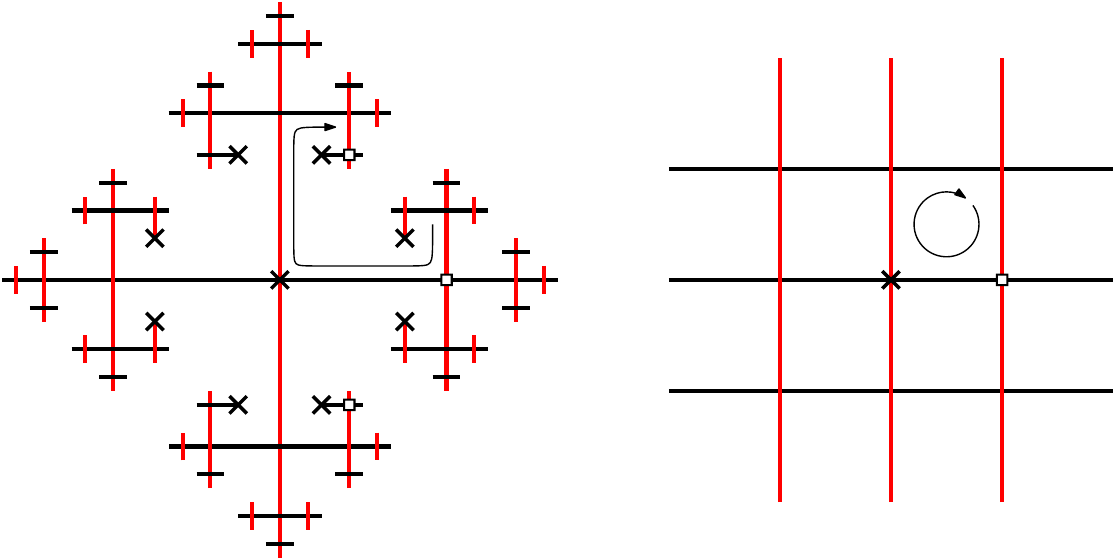}
		\par\end{centering}
	\caption{\label{fig:OhHow}The tree structure for square fishnet models (left) can be depicted as a square fishnet(right). Horizontal lines represent one field species of field in the theory, vertical lines represent the other, we can consistently consider them all oriented rightwards and upwards. The nodes of the tree we have marked with the same symbol are represented by the same fishnet vertex. Wrapping once a tile of the fishnet corresponds to jumping from a branch of the tree to the consecutive one.}
\end{figure}

\begin{figure}[p]
	\begin{centering}
		\includegraphics[width=\textwidth]{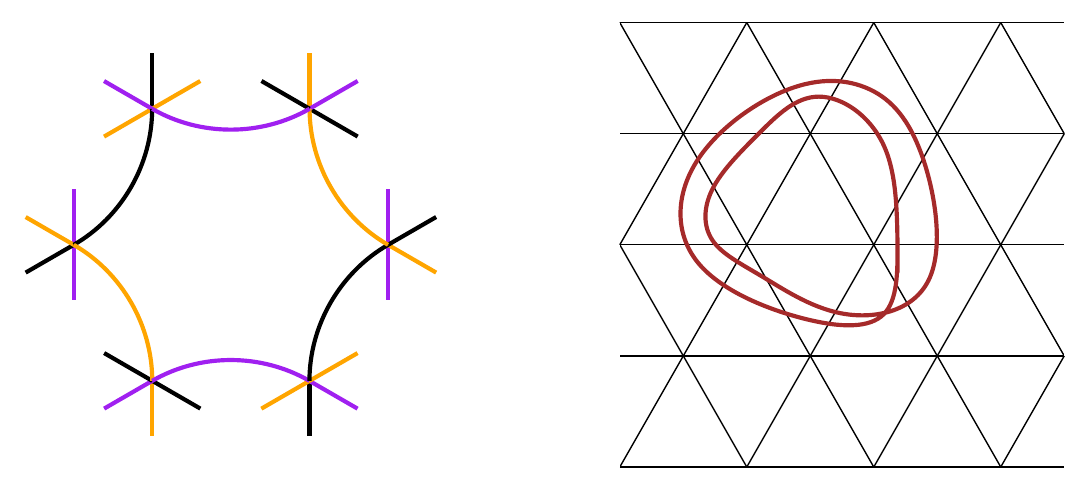}
		\par\end{centering}
	\caption{\label{fig:HexagonTile}The hexagon diagram in the 3 dimensional fishnet model and the path on the lattice that produces it. External legs could be contracted with each other or with other tiles. Different fields in the diagram are painted with different colors. In the lattice, the type of filed is uniquely determined by the direction of the line.}
\end{figure}

\begin{figure}[p]
	\begin{centering}
		\includegraphics[width=\textwidth]{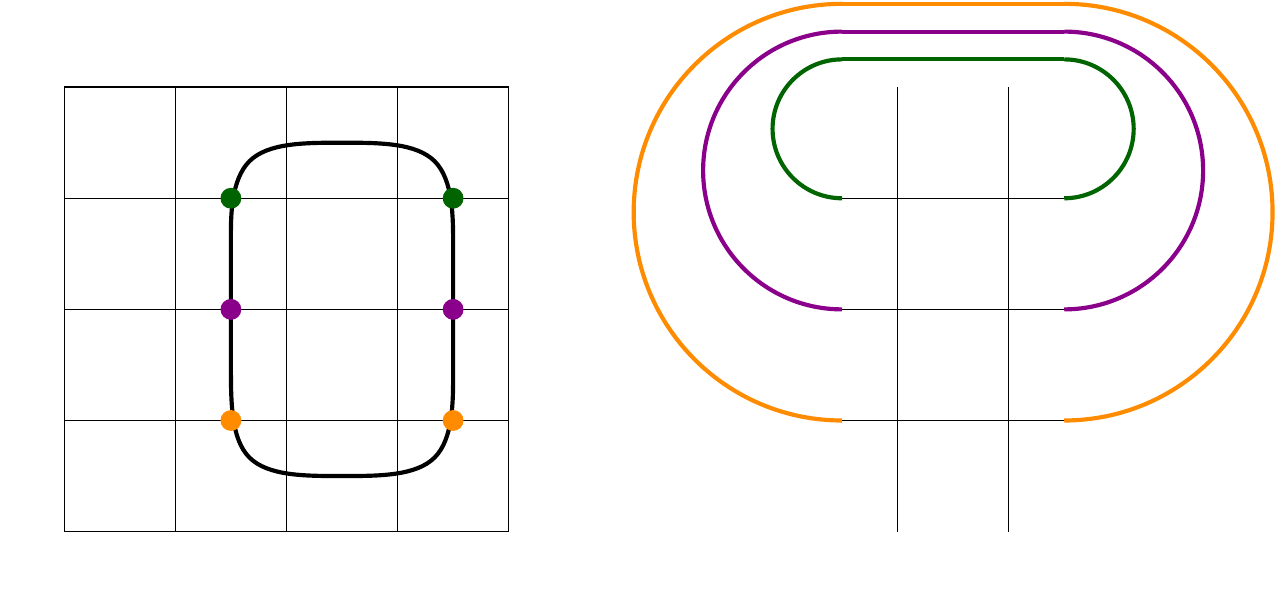}
		\par\end{centering}
	\caption{\label{fig:PathInLattice}An example on how to transform a path in a lattice into a diagram. Points coloured
with the same color are contracted afterwards. In the lattice, horizontal lines are $\phi_1$, vertical lines are $\phi_2$.}
\end{figure}

As an aside, let us mention that the graphic depictions of the tree diagram usually become impractical as their level increases due to the power growth in the number of legs and nodes at each level. It is convenient for our purposes to use a fishnet diagram to represent them. In this picture each vertex of the fishnet will represent different overlapped nodes of the tree in such a way that all overlapping legs are of the same type of field and the color structure is preserved in every vertex\footnote{Since there is only a single vertex in the theory that can be added to every external line, we can construct the minimal fishnet that serves this purpose using the following procedure: For each external line, we add vertices to its closest neighbours in each side until the two closest legs are antifields of the original external leg. In the fishnet picture, the next vertex added to these neighbouring legs will overlap with the vertex from which the original leg emanates, so that the three legs of the same field are drawn on top of each other with consistent orientation.}. Then, going once around a tile corresponds to jumping from a branch of the tree to the consecutive one. Figure \ref{fig:OhHow} illustrates this construction for the square fishnet model.
When we use this representation to depict the construction of generic planar diagrams, we notice that the single trace structures obtained after performing the simplest contractions are represented as closed paths (plausibly winding several times around different sets of tiles, see figure \ref{fig:HexagonTile}), on which we can perform additional contractions (see figure \ref{fig:PathInLattice}).
\afterpage{\clearpage}

\begin{figure}
\begin{centering}
\includegraphics[height=0.25\paperwidth]{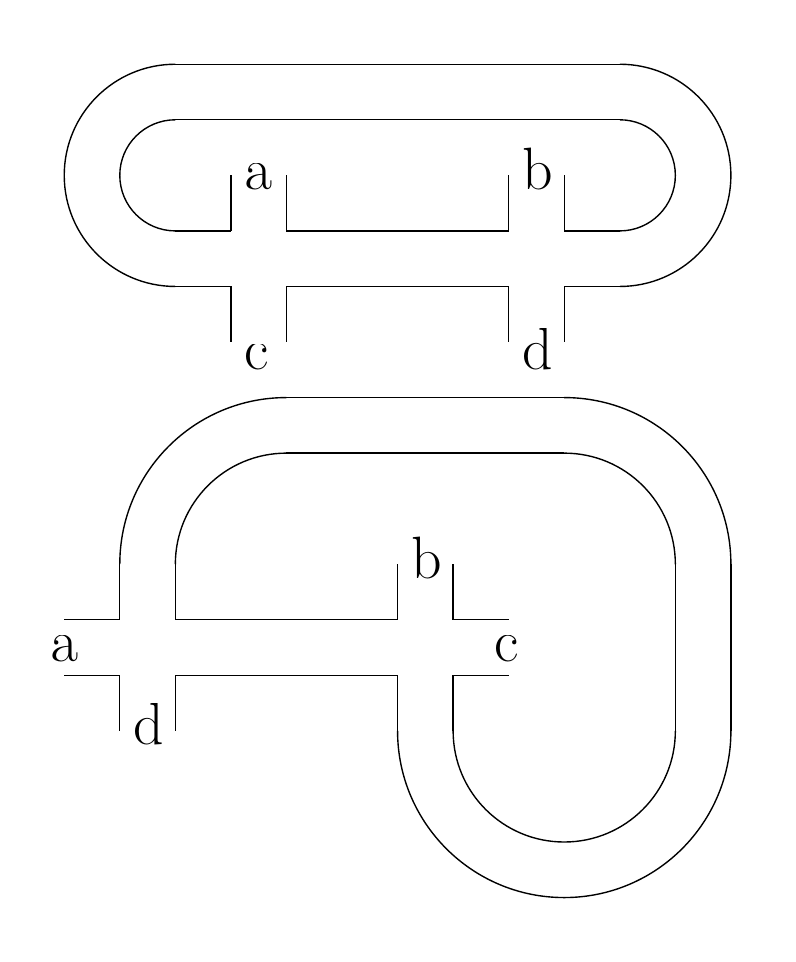}\includegraphics[height=0.25\paperwidth]{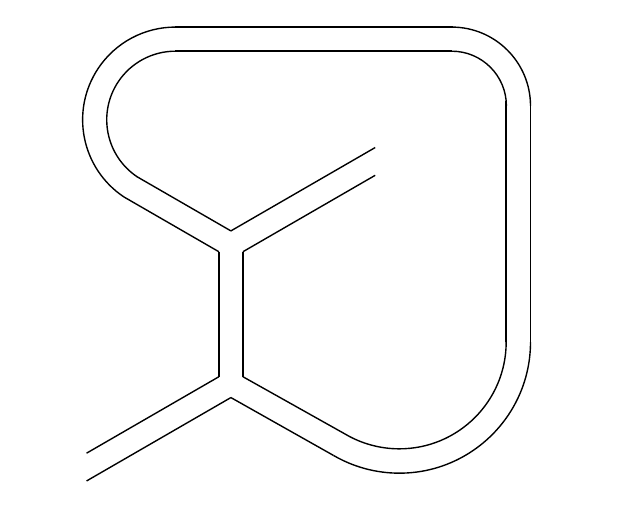}
\par\end{centering}
\caption{\label{fig:Divergences}Candidate planar divergences in the square
lattice models (left) and single trace divergence
arising in the hexagonal lattice model (right).}
\end{figure}

Equipped with our understanding of planar diagrams in fishnet models, we are now in position to discuss which among them can give rise to divergences. The simplest setting for addressing this question is dimensional regularization, in which contractions between a vertex and itself can be consistently put to zero.
Let us examine each type of fishnet separately:
\begin{itemize}
\item Triangular lattice: As we go along a circuit, each external leg is separated from any leg that could potentially be contracted with it by at least four other legs. Notice that we exclude the possibility of self-contractions in this counting. No additional contraction can reduce the amount of legs between the contracted vertices below 4. Therefore the minimal amount of external legs on each side of the contraction on the path is 4, and there are at least 8 external legs for each diagram with loops. As a result, the 3 dimensional theory introduced by \cite{Caetano:2016ydc} is conformal, at least perturbatively, in the planar limit.
\item Hexagonal lattice: The minimal amount of external legs on each side of a contraction of the path is 1. However, traces of single fields vanish in semisimple theories when the Gauss' law is imposed to external states. As a result, for the scalar theory in 6 dimensions there is no divergent diagram.
\item Square lattice: The minimal amount of external legs on each side of a contraction of the path is 2, and indeed the presence of loops with four external lines radiates double trace operators of the field squared in all theories of this fishnet type.
\end{itemize}
\afterpage{\clearpage}
The list of potentially divergent diagrams in each of the theories of \eqref{eq:S1ProposedLagrangians} is summarized in table \ref{tab:Divergences}. Figure \ref{fig:Divergences} illustrates the simplest forms in which $\text{tr}^2\phi$ and $\text{tr}^2\phi^2$ divergences can appear.

\begin{table}
\begin{centering}
\begin{tabular}{|c|c|c|c|c|}
\hline 
Vertex & Lattice & D & Div. & Pathology\tabularnewline
\hline 
\hline 
$\text{tr}\phi^{6}$ & Tri. & 3 & $3-\frac{1}{2}e_{\phi}$ & $\emptyset$\tabularnewline
\hline 
$\text{tr}\phi^{4}$ & Sq. & 4 & $4-e_{\phi}$ & $\text{tr}^{2}\phi^2$\tabularnewline
\hline 
$\text{tr}\phi^{2}\psi^{2}$ & Sq. & 3 & $3-\frac{1}{2}e_{\phi}-e_{\psi}$ & $\text{tr}^{2}\phi^2$, $\text{tr}^2\left(\phi\psi\right)^2$, $\text{tr}^2 \phi^3$ \tabularnewline
\hline 
$\text{tr}\psi^{4}$ & Sq. & 2 & $2-\frac{1}{2}e_{\psi}$ & $\text{tr}^{2}\psi^2$\tabularnewline
\hline 
$\text{tr}\phi^{3}$ & Hex. & 6 & $6-2e_{\phi}$ & ($\text{tr}^2\phi$)\tabularnewline
\hline 
$\text{tr}\phi\psi^{2}$ & Hex. & 4 & $4-e_{\phi}-\frac{3}{2}e_{\psi}$ & ($\text{tr}^2\phi$), $\text{tr}^{2}\phi^2$\tabularnewline
\hline 
\end{tabular}
\par\end{centering}
\caption{\label{tab:Divergences}For each theory in \ref{eq:S1ProposedLagrangians} here the schematic form of the vertex, its dimension, the superficial degree of divergence of its diagrams with $e_{\phi}$ external bosons and $e_{\psi}$ external fermions. The last column lists the divergent diagrams present in the perturbative expansion. Since none of the divergences in the list can be absorbed by the counterterms of the Lagrangian, only semisimple $\phi^{3}$ theories (when the Gauss's law is imposed on external states) and
$\phi^{6}$ theories are immune to the radiation of additional terms in the Lagrangian.}
\end{table}

\section{\label{sec:Double traces}Double traces for fishnet models}

Double trace operators are a central piece of the puzzle of integrability-preserving deformations
of $\mathcal{N}=4$ SYM and ABJM in absence of supersymmetry. The stabilization of their RG flow
is the main difficulty in the construction of these theories. In this
section we will briefly comment on the implications of their presence
in models that would otherwise be of the ``fishnet'' type. Additionally,
we will introduce a type of deformation for fishnet candidate Lagrangians,
refinement, which protects them from developing double
trace contributions.

\subsection{\label{subsec:DT}Fishnets within double trace corrected theories}

Most of the theories analysed in the previous section develop double
trace operators under RG flow which break down the fishnet structure
and in general render the model non-conformal. Indeed, both in the
context of orbifolds \cite{Bershadsky:1998mb, Bershadsky:1998cb}
and real gamma deformations \cite{Fokken:2013aea}, these terms rarely allow for perturbatively reachable stable
fixed points: the zeroes of their one-loop beta functions generically lie in the
imaginary axis of the complex plane. In fact, in a wide class of orbifolds, a no-go theorem precludes the existence of fixed points in nonsupersymmetric theories \cite{Dymarsky:2005nc, Dymarsky:2005uh}.

When we relax the unitarity constraint, however, the restrictions
on the availability of fixed points will not come from reality anymore.
It is, to the knowledge of the authors, unclear whether additional
constraints on the complexified parameter space will in general arise
from demanding a consistent nonperturbative completion of the theory.
However, at a speculative level, it was proposed in \cite{Sieg:2016vap}
to consider the scenario where appropriate fine-tuning of the complexified
couplings of double trace terms allows us to perturbatively access
a true stable fixed point of the theory under RG. Without committing
ourselves to any conclusion on the viability of this construction,
we will in what follows briefly analyse what observables in the theory
are sensitive to the presence of double trace operators, and identify
the sector left invariant by their presence.

\begin{figure}
\begin{centering}
\includegraphics[width=\textwidth]{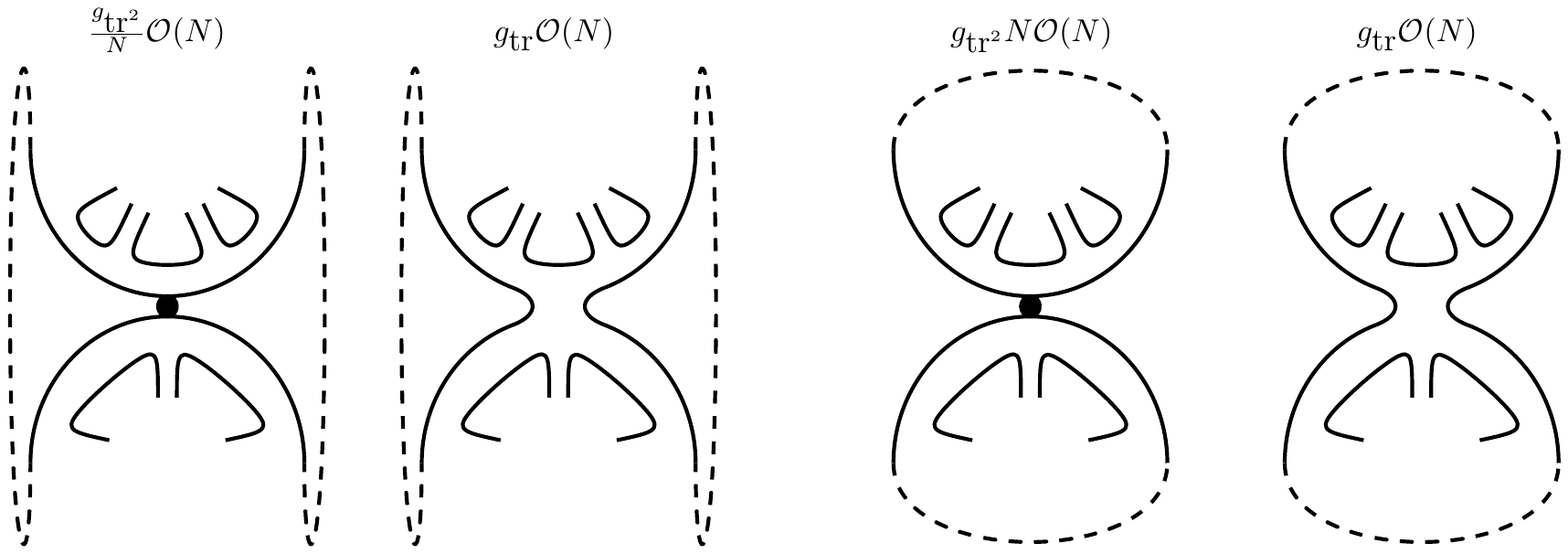}
\par\end{centering}
\caption{\label{fig:BalconiesToBridge}We depict in the double line notation
the effect of replacing a generic double trace operator by a corresponding
single trace construction, plausibly made out of several vertices. In the process the amount of loops either increases or decreases by a unit.}
\end{figure}

A simple way to discuss the large $N$ behavior of diagrams with at
least a double trace insertion it to observe what happens when this
double trace insertion is replaced by a single trace vertex with the
same external legs, see figure \ref{fig:BalconiesToBridge}.
In regard to the 't Hooft limit counting it is not even necessary
that such a single trace vertex is part of the theory. In the double
line notation terminology, depending on whether the two line segments
that get rearranged in the process were part or not of the same line,
a loop is created or annihilated. Consequently, any double trace coupling
with a vanishing $\beta$ function in the 't Hooft limit will appear
suppressed by a $N^{-1}$ factor in the action, but contribute at the planar level
to any diagram where the double trace insertion splits the color structure
in two disconnected planar components.This $N^{-1}$ factor coincides
with the one appearing in the 1-loop analysis of \cite{Sieg:2016vap}.

The functional formalism for field theory offers a particularly convenient
way to construct a generating functional for the observables that
remain unaffected by double traces at the planar level. From our previous
discussion it should be clear that if double trace operators are the
only possible cause for the splitting of the color double line diagram
in disjoint components, they can only contribute at planar level when
each of the double trace insertions causes an additional factorization
of this type. Therefore, a Legendre transform with respect to bi-local
sources, each one having its two positions coupled to the two traces
in a specific double trace operator, will behave as the generating
functional for the irreducible diagrams that stand protected from
double trace planar corrections.

\subsection{\label{subsec:NoDT}Theories protected from double traces by additional flavour}

In the analysis of the perturbative sources of double trace operators
performed in section \ref{sec:Divergs} it became manifest
that the addition
of more vertices to a planar diagram would necessarily lead to an
increase of the number of external legs. It is possible to take advantage
of this fact when building models in order to completely deprive the double trace
operators from entering the planar sector.

The simplest way to implement this is by a super-selection process that we shall denominate \emph{refinement}:
we add to each of the original species in the proposed action $\mathcal{S}^{\left(\text{proposed}\right)}$ up to two flavour indices in a specific manner that do not allow
a simultaneous diagonalization of the free and interacting part of
the Lagrangian. More precisely, we will choose in the diagonal basis
for the free Lagrangian, an interaction that cyclically transits between
flavours, 
\[
\mathcal{L}_{i,j}^{\left(int\right)}\propto\delta_{i,j\text{mod}n+1};\:i,j\in\left\{ 1,\ldots,n\right\} 
\]
In practice, this choice acts as a super-selection rule on the Feynman
diagrams, that, provided the cycles are sufficiently large projects
out all superficially divergent diagrams that could radiatively source
double trace terms.

\begin{figure}
\begin{centering}
\includegraphics[width=0.6\textwidth]{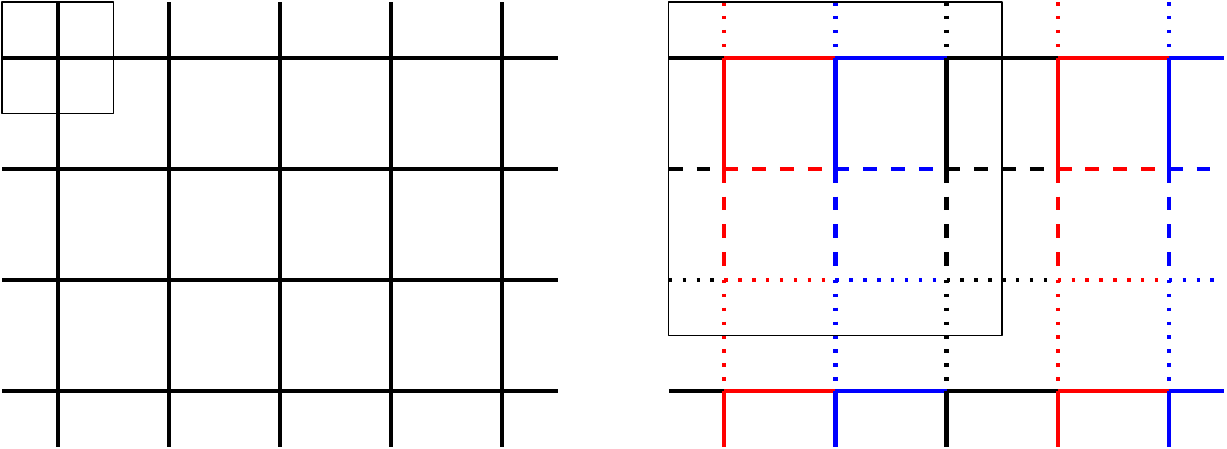}
\par\end{centering}
\caption{\label{fig:GuKa}The refinement of the G\"urdo\u{g}an-Kazakov's model with
two flavour indices $a,b\in\left\{ 1,2,3\right\} $ super-selects
from the fishnets of the original proposal (left) only those compatible
with the resulting $3\times3$ color structure (right). The distinction
is important when we study the diagrams which wrap $\emptyset$ a cylinder,
such as those that compute anomalous dimensions of $\text{tr}\phi^{L}$ The minimal periodic structure that allows to put the construction on the cylinder is in each case the unitary cell enclosed by a thin line.}
\end{figure}

This strategy can be applied to the G\"urdo\u{g}an-Kazakov square scalar
fishnet candidate \cite{Gurdogan:2015csr}, as figure \ref{fig:GuKa}
 illustrates. The resulting theory has three-indexed constituent scalar fields $\phi_{i,a,b}$. One of the indices, $i\in\left\{1,2\right\}$, distinguishes their role in the fishnet structure; the other two, $a,b\in\left\{1,2,3\right\}$, are the indices introduced to refine the theory. The Lagrangian of the refined model reads

\begin{equation}
\begin{array}{ccl}
& & \mathcal{S}_{+}^{\left(\text{refined}\right)}=
N\int d^{4}x\text{Tr} \left( \frac{1}{2}\sum_{i=1,a,b=1}^{2,3}\left|\partial_{\mu}\phi^{i}_{a,b}\right|^{2}+\mathcal{L}_{\text{int}}\right) 
\\
\mathcal{L}_{\text{int}}&=&\sum_{a,b=1}^{3}\xi_{a,b}\phi_{1,a,b}\phi_{2,a,b}\phi_{1,a_{\text{mod}3}+1,b}^{\dagger}\phi_{2,a,b_{\text{mod}3}+1}^{\dagger}=\\
&=&\sum_{a_1,a_2,b_1,b_2=1}^{3}\xi_{a_1,b_1}\phi_{1,a_1,b_1}\phi_{2,a_1,b_1}\phi_{1,a_2,b_1}^{\dagger}\phi_{2,a_1,b_2}^{\dagger}\delta_{a_1,a_2\text{mod}3+1}\delta_{b_1,b_2\text{mod}3+1}
\end{array}
\end{equation}

\section{\label{sec:AllOrders}Anomalous dimensions in the hexagonal fishnet model}

The diagrammatic simplicity of the fishnet models reveals itself in its full magnificence when we address the computation of anomalous dimensions of single trace operators. For them, the only allowed planar diagrams can be obtained by compactifying a 
periodic direction of the fishnet on a circle, plausibly with some shift, i.e., mixing the two translational symmetries of the unitary cell of the fishnet lattice. Not any cylindrical
construction will be compatible with a given single trace operator,
though: Both the shift and the size of the compact direction $L$ in periodic
unitary cells will be fixed uniquely by the given external operator. In contrast,
the number of cells along the cylinder is not fixed and
these operators develop a nontrivial dependence on the coupling. 

The examples that conclude the present paper, namely the anomalous dimensions for the operators of $L=2$ in this model ($\mbox{tr}\phi_i \phi_j$ operators) to all orders, illustrate simultaneously the simplicity and nontriviality of the hexagonal fishnet model. For convenience, let us evoke the action of this theory, which was already presented in section \ref{sec:Divergs}:
\begin{equation}
\mathcal{S}_{\Yup\Ydown}=N\int d^{D}x\text{Tr}\left(\frac{1}{2}\sum_{i=1}^{3}\left|\partial_{\mu}\phi^{i}\right|^{2}+\xi_{1}\phi_{1}^{\dagger}\phi_{2}\phi_{3}+\xi_{2}\phi_{1}\phi_{2}^{\dagger}\phi_{3}^{\dagger}\right)
\end{equation}
Here, $D$ will be taken to be $6-2\epsilon$ in the setting of dimensional regularization. Notice that in this model the two vertices will appear alternately in any diagram. In practice, this implies that the perturbative expansion for any observable can be arranged in terms of the quantity
\begin{equation}
\alpha\equiv\frac{\xi_1 \xi_2}{\left(4\pi\right)^{D/2}}
\end{equation}

The anomalous dimensions of $\mbox{tr}\phi_i \phi_j$ operators in the hexagonal fishnet model can be computed at all perturbative orders via the resolution of a closed integral equation for a self-energy. 
The perturbative expansion up to an arbitrary order $n$ in $\alpha$ is also possible: the problem amounts to calculating the first $n$ terms in the series expansion of a known analytic function. The former method will be presented in this section. The comparison to the results of the latter method is described in detail in the appendix.

\subsection{\label{subsec:vacuum}Anomalous dimension of $\text{tr}\phi_1^2$}
Let us first address the computation of the anomalous dimension of
$\text{tr}\phi_{1}^{2}$. Our theory is conformal and the operator
under consideration does not undergo any mixing, so in dimensional
regularization (with $D=6-2\epsilon$) the relations between the bare
and renormalized quantities of interest are
\begin{equation}
\begin{array}{c}
\alpha_{0}=\mu^{2\epsilon}\alpha\rightarrow\beta_{\alpha}=-2\epsilon\alpha\\
\left(\text{tr}\phi_{1}^{2}\right)_{0}=\mathcal{Z}_{11}\left(\alpha\right)\text{tr}\phi_{1}^{2};\;\gamma_{11}=\frac{d\log\mathcal{Z}_{11}}{d\log\mu}=2\epsilon\alpha\frac{d\log\mathcal{Z}_{11}^{-1}}{d\alpha}
\end{array}\label{eq:RGFlow}
\end{equation}
A particular trait of our theory is the absence of field renormalization
for the $\phi_{i}$ fields at the planar level. This allows us to
read directly the anomalous dimension $\gamma_{11}$ from the renormalized
correlator
\begin{equation}
\Sigma_{11}\left(p\right)\equiv\left\langle \phi_{1}^{\dagger}\left(p\right)\phi_{1}^{\dagger}\left(-p\right)\text{tr}\phi_{1}^{2}\left(0\right)\right\rangle _{1PI}=\mathcal{Z}_{11}^{-1}\cdot\left(\Sigma_{11}\right)_{0}\label{eq:TheSigma_1}
\end{equation}
where the subindex 1PI indicates that the correlator is 1 particle irreducible (with amputated external legs). From this point on, the conventional approach in perturbative analysis
is to series expand $\left(\Sigma_{11}\right)_{0}$ in $\alpha$ and
choose a prescription for the expansion
\begin{equation}
\mathcal{Z}_{11}^{-1}=1+\sum_{n=1}^{\infty}\alpha^{n}z_{n}=1+\sum_{n=1}^{\infty}\alpha^{n}\sum_{m=0}^{n}\epsilon^{-m}a_{n,m}\label{eq:ExpandZ}
\end{equation}
that renders the quantity $\Sigma_{11}$ finite. However, this particular
problem permits a much more powerful approach that provides us with
the re-summed quantity $\mathcal{Z}_{11}^{-1}$ to all perturbative
orders. For this reason, the conventional perturbative analysis of
this quantity has been relegated to the appendix, and in what follows
we explain how this re-summed result is obtained. The concordance
of the results of the two appraoches provides us with a nontrivial
sanity check for the result.

\begin{figure}
\begin{centering}
\includegraphics[width=1\textwidth]{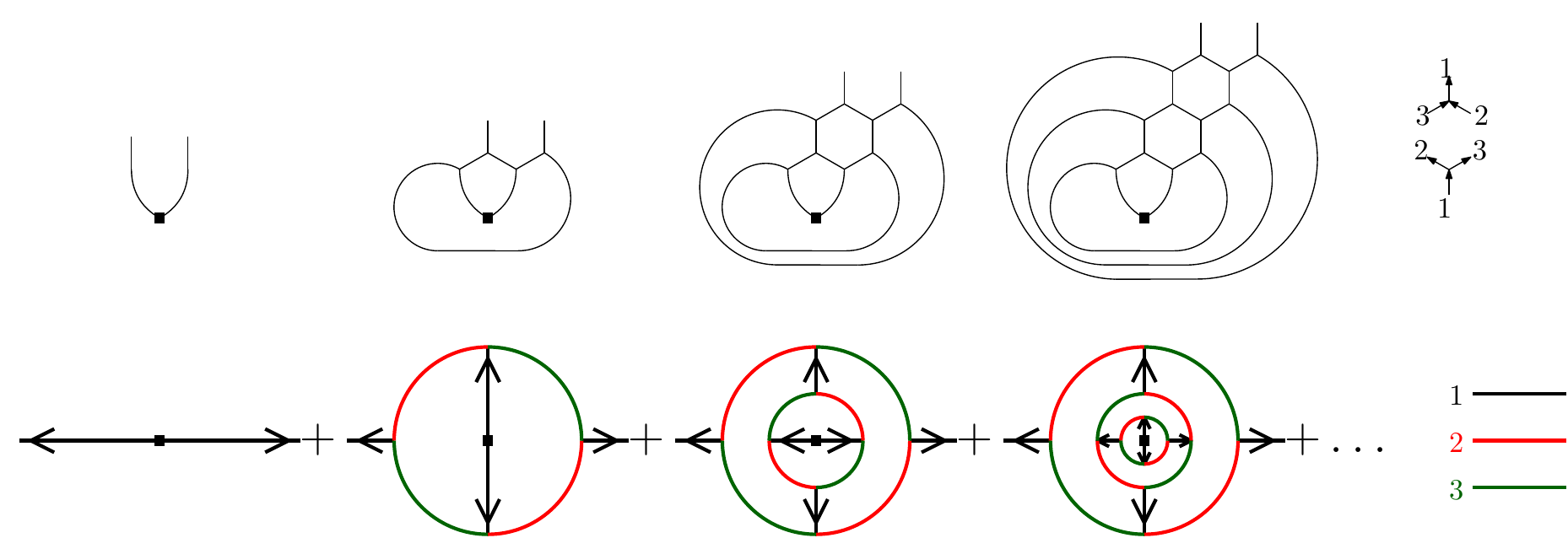}
\par\end{centering}
\caption{\label{fig:Diagram}Diagrammatic expansion of $\Sigma_{11}$. Each color represents a different type of field, according to the legend on the right. The missing arrows can be inferred from the chirality of the vertices, they have been suppressed for the sake of clarity. The alternative representation of each diagram depicted above makes the fishnet structure explicit by associating a direction on the plane to each field.}
\end{figure}

The key point in the aforementioned re-summation is to notice that
the perturbative expansion of $\Sigma_{11}$ has a nested structure,
as depicted in figure \ref{fig:Diagram}. 
This allows us to write
the following integral equation:
\begin{equation}
\Sigma_{11}\left(p\right)=\mathcal{Z}_{11}^{-1}+\alpha^{2}\pi^{2\epsilon-6}\mu^{4\epsilon}\int\frac{d^{6-2\epsilon}qd^{6-2\epsilon}r}{q^{2}\left(q+p\right)^{2}r^{2}\left(r+p\right)^{2}\left(q-r\right)^{4}}\Sigma_{11}\left(q-r\right)\label{eq:SD1_1}
\end{equation}
The terms in the right hand side of this equation have divergences
at $D\rightarrow6$ that cancel among themselves. It is possible to
sidestep this difficulty by applying the Laplacian operator to this
equation. The resulting expression contains no divergence and therefore
we can evaluate it at $\epsilon\rightarrow0$:
\begin{equation}
\nabla^{2}\Sigma_{11}\left(p\right)=-4\alpha^{2}\pi^{-6}\int\frac{d^{6}qd^{6}r}{q^{2}\left(q+p\right)^{4}r^{2}\left(r+p\right)^{4}\left(q-r\right)^{2}}\Sigma_{11}\left(q-r\right)\label{eq:SD2_1}
\end{equation}

\begin{figure}
	\begin{centering}
		\includegraphics[width=1\textwidth]{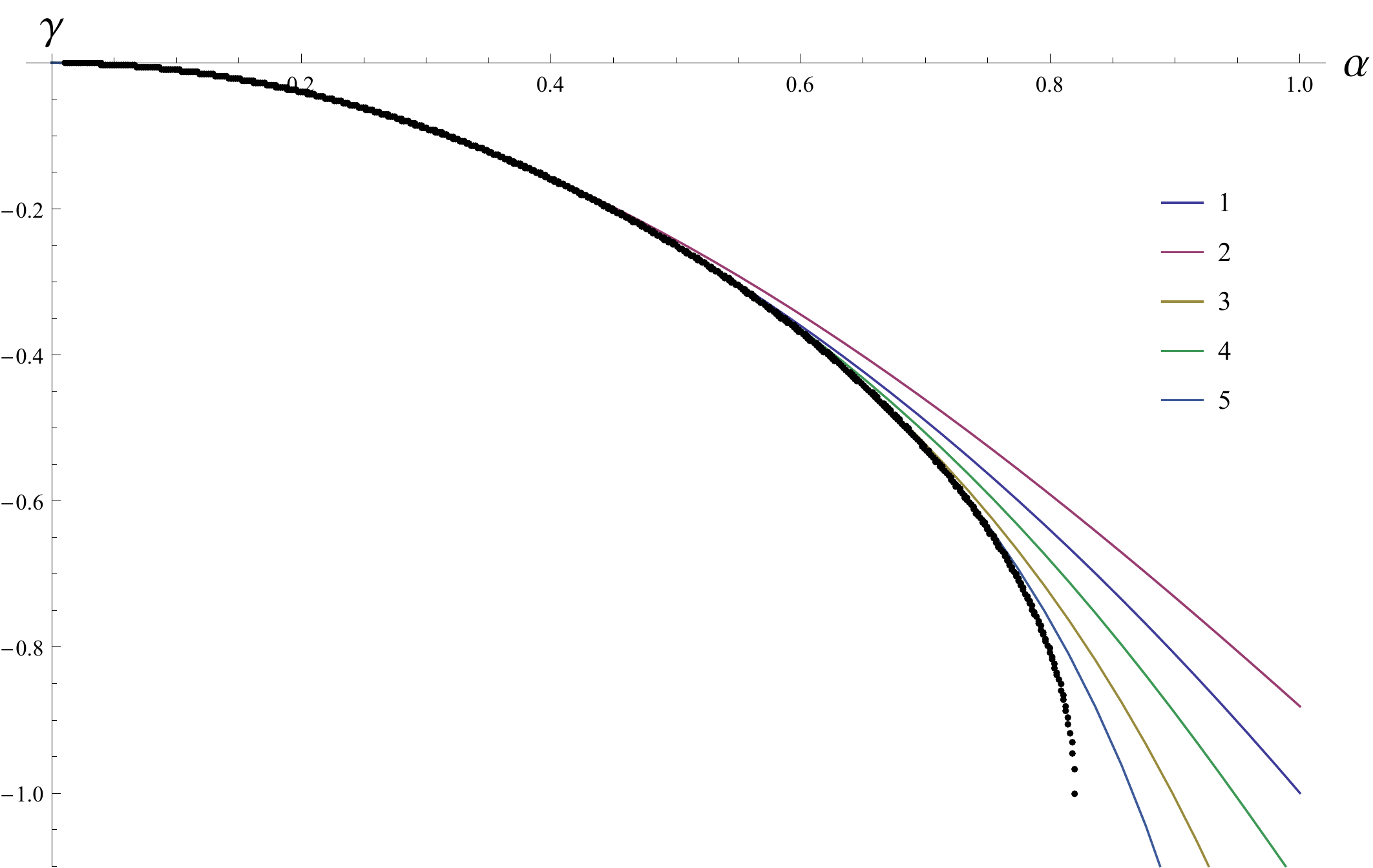}
		\par\end{centering}
	\caption{\label{fig:ConvergencePhiSquared}Convergence of the perturbative expansion to the exact result for the anomalous dimension $\gamma_{11}$. The thin lines correspond to truncated sums of the perturbative expansion at the orders indicated in the legend. The thick black dots are numerical evaluations of the re-summed result.}
\end{figure}

Because $\Sigma$ is three point correlator of the conformal field
theory when one of its momenta is put to zero, it can be shown it
is of the form 

\begin{equation}
\Sigma_{11}=\sigma\left(\frac{\mu^{2}}{p^{2}}\right)^{\xi}\label{eq:Ansatz_1}
\end{equation}
where $\sigma$ denotes a constant that plausibly depends on $\alpha$.
The relation between the anomalous dimension we want to compute and
$\xi$ can be directly inferred from dimensional analysis in the Fourier
transform of the three point function amplitude:
\begin{equation}
\begin{array}{c}
-\mbox{Power of p in \ensuremath{\Sigma}}=\mbox{\# integrals}\cdot D-\sum\Delta_{\mbox{free}}-\sum\gamma-\Delta_{\mbox{amputation}}\\
\xi=\frac{2D}{2}-\frac{4}{2}\Delta_{\phi}-\frac{1}{2}\gamma_{11}-\frac{2}{2}\Delta_{\phi}=-\frac{1}{2}\gamma_{11}
\end{array}\label{eq:NaiveCounting}
\end{equation}
All in all, we obtain a closed transcendental equation for $\gamma_{11}$:
\begin{equation}
\gamma_{11}\left(\gamma_{11}+4\right)=-4\alpha^{2}\pi^{-6}\int\frac{d^{6}qd^{6}r}{q^{2}\left(q+p\right)^{4}r^{2}\left(r+p\right)^{4}\left(q-r\right)^{2-\gamma_{11}}}=-4\alpha^{2}I_{2,2,1-\gamma_{11}/2}\label{eq:Transcendental}
\end{equation}
The function $I_{2,2,1-\gamma_{11}/2}$ can be computed using an appropriate expansion in Gegenbauer polynomials \cite{Kotikov:1995cw} or inferred from recurrence relations 
\cite{Broadhurst:1996ur}. In the notation of \cite{Broadhurst:1996ur}, \renewcommand{\arraystretch}{1.3}
\begin{equation}
\begin{array}{c}
I_{a,b,c}\equiv I_{a,b,c,a+b+c-D/2}\equiv \pi^{-6}\int\frac{d^{6}qd^{6}r}{q^{2}\left(q+p\right)^{2a}r^{2}\left(r+p\right)^{2b}\left(q-r\right)^{2c}}\\
\left(D-3\right)I_{a,b,c,d}=b\,d\,G_{1,d+1}\left(G_{a,c+1}S_{D/2-a-1,b-1,D/2+a-d-2,d-b}+\{a\leftrightarrow b\}\right)\\
G_{x,y}=\Gamma(D/2-x)\Gamma(D/2-y)\Gamma(x+y-D/2)/\left(\Gamma(x)\Gamma(y)\Gamma(D-x-y)\right)\\
S_{a,b,c,d}=\pi\cot\left(\pi c\right)\left(H_{a,b,c,d}\right)^{-1}-c^{-1}-\left(c^{-1}+d^{-1}\right)F_{a+c,-b,-c,b+d}\\
H_{a_{1},a_{2},a_{3},a_{4}}=\Gamma(1+a+b+c+d)\prod_{i=1}^{4}\Gamma(1+a_{i})/\prod_{i,j=1}^{2}\Gamma(1+a_{i}+a_{2+j})\\
F_{a,b,c,d}=\pFq{3}{2}{-a,-b,1}{1+c,1+d}{1}-1\;.
\end{array}\label{eq:TheI}
\end{equation}
\renewcommand{\arraystretch}{1.1}Equation (\ref{eq:Transcendental})
can be resolved numerically, picking among the different branches
of the solution the one that has the appropriate classical limit at
$\alpha\rightarrow0$. Figure \ref{fig:ConvergencePhiSquared} displays the
results of this computation and compares it to the predictions of
the leading perturbative terms computed in the appendix.

Notice the abrupt ending in a vertical slope of the physical branch of solutions to (\ref{eq:Transcendental}) at the points in which the anomalous dimension becomes $-1$, namely at
\begin{equation}
\alpha=\pm \frac{3}{4}\sqrt{\frac{3}{19-18C}}
\end{equation}
where C denotes the Catalan constant. At this point, this real branch of solutions folds back into another branch of the solution. The roots corresponding to these two branches become complex beyond this point of coalescence, and there is a priori no reason to identify one of them as the physically relevant one. As we discussed in the introduction, it is tempting to look for a physical interpretation of these breakdowns, but a meaningful discussion requires a better understanding of the complete spectrum or a nonperturbative formulation of the theory.

%

\subsection{\label{subsec:Magnon}Anomalous dimension of $\mbox{tr}\phi_1\phi_2$}
\begin{figure}
	\begin{centering}
		\includegraphics[width=1\textwidth]{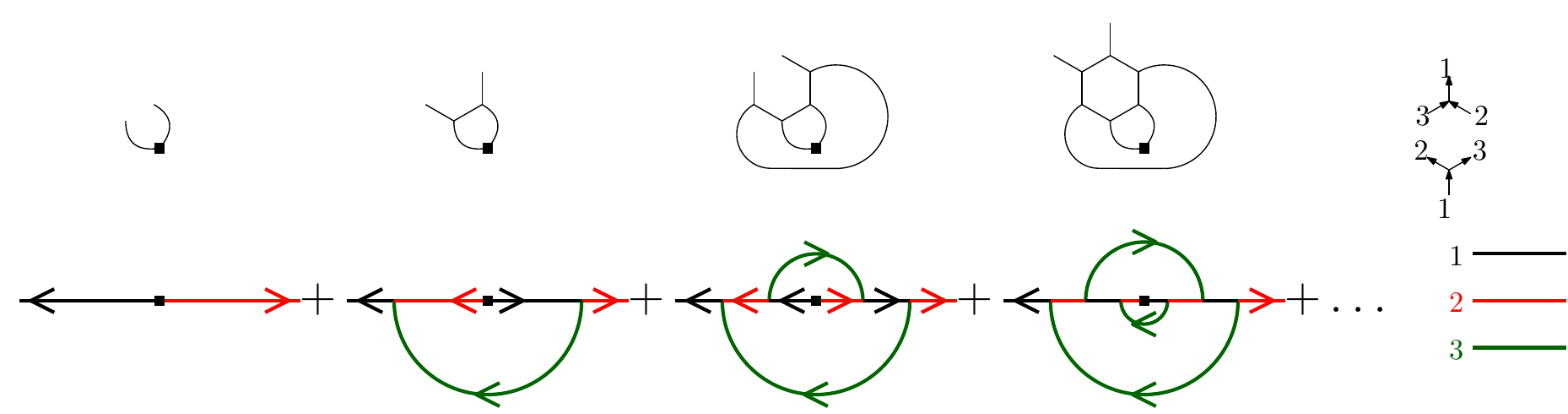}
		\par\end{centering}
	\caption{\label{fig:Diagram2}Diagramatic expansion of $\Sigma_{12}$. The conventions are identical to those of figure \ref{fig:Diagram}.}
\end{figure}

The anomalous dimension of $\mbox{tr}\phi_1\phi_2$ can be obtained using a procedure that, in form, is identical to the one we presented for $\mbox{tr}\phi_1^2$. The diagrams are also organized in a nested structure, and all the discrepancies between the two cases are ultimately reduced to the different shape of the basic building block in this sequence, made explicit by the comparison of figure \ref{fig:Diagram2} to  figure \ref{fig:Diagram}.

The quantity of interest in this case is
\begin{equation}
\Sigma_{12}\left(p\right)\equiv\left\langle \phi_{1}^{\dagger}\left(p\right)\phi_{2}^{\dagger}\left(-p\right)\text{tr}\phi_{1}\phi_2\left(0\right)\right\rangle _{1PI}=\mathcal{Z}_{12}^{-1}\cdot\left(\Sigma_{12}\right)_{0}\label{eq:TheSigma_2}
\end{equation}
By following the steps used in the computation of $\Sigma_{11}$ we now obtain
\begin{equation}
\nabla^2\Sigma_{12}\left(p\right) = -4 \alpha \pi^{-3}
\int \frac{d^6 q }{q^{4} \left(p+q\right)^{4}} \Sigma_{12}\left(q\right)\;.
\label{eq:SD2_2}
\end{equation}
and again, we use the Ansatz
\begin{equation}
\Sigma_{12}=\tilde{\sigma}\left(\frac{\mu^2}{p^2}\right)^{\xi}
\end{equation}
with $\xi=-\frac{1}{2}\gamma_{12}$.

The solution here is much simpler than in the previous case, since the equation for the exponent becomes algebraic:
\begin{equation}
\gamma (\gamma+4)=-4\alpha
\frac
{\Gamma \left(1\right) \Gamma \left(1-\frac{\gamma}{2}\right)    \Gamma \left(1+\frac{\gamma}{2}\right)}
{\Gamma \left(2\right) \Gamma \left(2+\frac{\gamma}{2}\right)    \Gamma \left(2-\frac{\gamma}{2}\right)}
=\frac{-4\alpha}{1-\left(\frac{\gamma}{2}\right)^2}
\end{equation}
Among the four branches of solutions to this equation, given by $\gamma_{12}=-1\pm \sqrt{5\pm 4 \sqrt{1+\alpha}}$, the one that reaches the classical value at $\alpha\rightarrow 0$ is
\begin{equation}
\gamma_{12}=-1+\sqrt{5-4\sqrt{1+\alpha}}
\end{equation}
These results can be compared to the perturbative computation described in appendix \ref{app:PerturbativePhi1Phi2}. 
\begin{figure}
	\begin{centering}
		\includegraphics[width=1\textwidth]{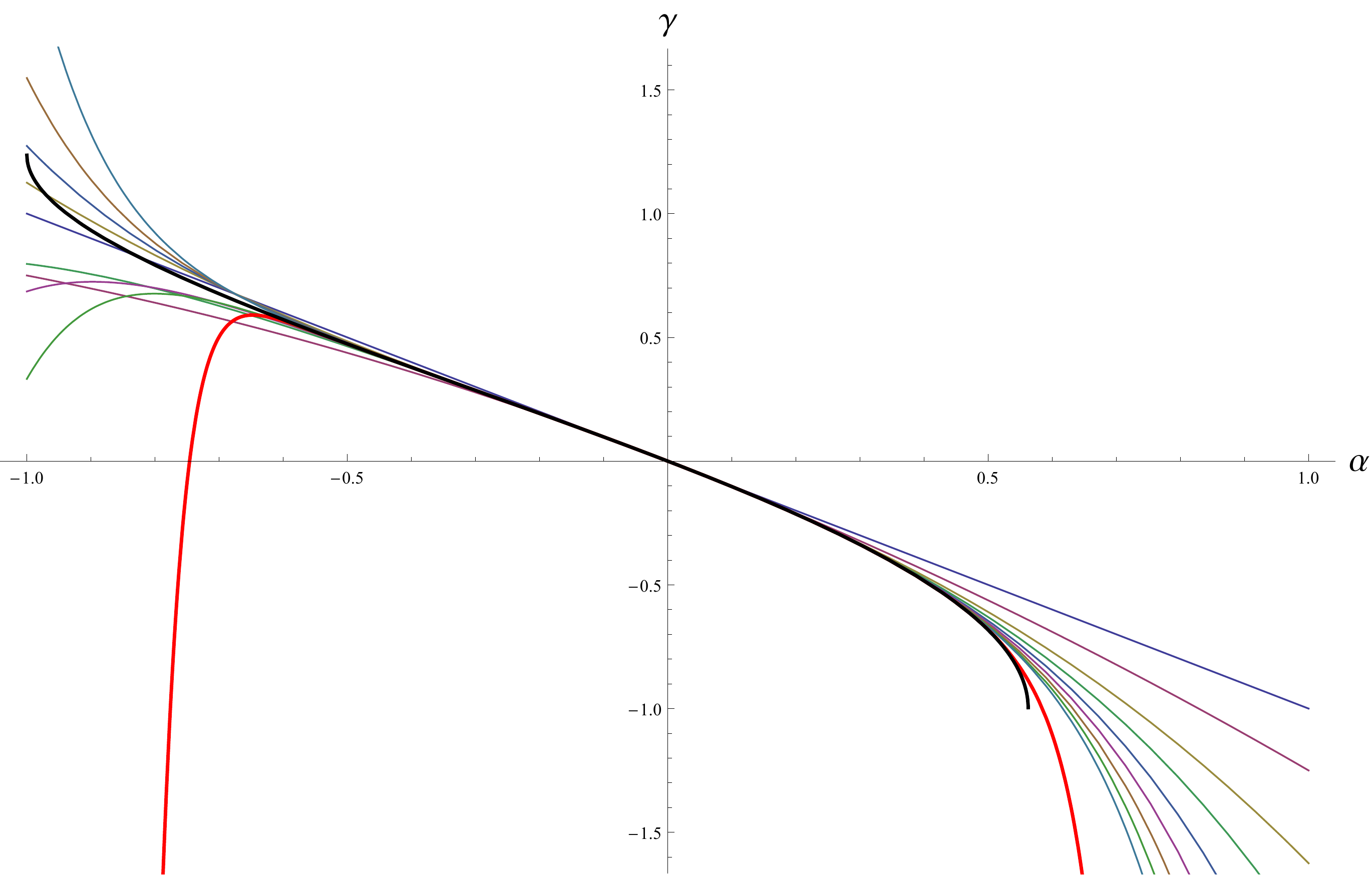}
		\par\end{centering}
	\caption{\label{fig:ConvergencePhi1Phi2}Convergence of the perturbative expansion to the exact result obtained for the anomalous dimension $\gamma_{12}$. The thin lines are truncated sums of the perturbative expansion at increasing orders. The red line is the 22nd order, the highest the authors deemed appropriate to calculate. The thick black line is the exact result. Notice that the segment $\alpha\in(-1,-9/16)$ lies manifestly outside the radius of convergence of the perturbative series.}
\end{figure}

In contrast to what happens for $\gamma_{11}$, the behaviour of the anomalous dimension $\gamma_{12}$ is not symmetric in $\alpha$. Instead, it decreases monotonically between $(\alpha,\gamma_{12})=(-1,\sqrt{5}-1)$ and $(\alpha,\gamma_{12})=(9/16,-1)$. Both endpoints correspond to collisions of two real branches of solutions that become complex beyond them.

\subsection{\label{subsec:Integrability}Integrability perspective}

As an epilogue to this chapter let us briefly comment on the interest of the present models from the perspective of integrability.

The organization of single trace two-point amplitudes in cylindrical diagrams of fixed radius allows for a translation of the problem into the language of 1+1 spin chains, in close analogy to what happens in $\mathcal{N}=4$ SYM or ABJM for the chiral sector. The discussion on the integrability of these spin chain models can be addressed from two different perspectives. On the one hand, some fishnet models emerge as gamma deformations of $\mathcal{N}=4$ SYM and ABJM, and therefore, at least these cases are expected to share the same integrability properties, despite the fishnet case presents some structural differences in relation to the undeformed case\footnote{Consider for instance, the operators with no shift in the cylinder, which are the fishnet correspondent to BMN vacua. In the fishnet theory, these observables are unprotected, and their anomalous dimension can be computed up to a finite order using the Y-system/Asymptotic Bethe Ansatz approach\cite{Gurdogan:2015csr}}. See \cite{Caetano:2016ydc} for a complete discussion on this integrability approach. On the other hand, as we mentioned in the introduction, Zamolodchikov proved in \cite{Zamolodchikov:1980mb} that in scalar regular fishnet models the planar diagrams satisfy a Yang-Baxter equation in terms of their position/momentum propagators, which opens the perspective of applying integrability techniques of noncompact representations of orthogonal groups (see \cite{Chicherin:2012yn} and references therein, for instance). The simultaneous availability of these two types of integrability (one of which is strictly proven) makes this type of models an unparalleled probing ground to test integrability and understand it better.

The all-order results presented in this section can be viewed as the simplest implementation of Zamolodchikov's integrability, in the sense that ultimately our equations are reduced to an eigenvalue problem for the dilatation operator. From the spin chain perspective, nevertheless, the amplitudes computed correspond to operators with wrapping effects \cite{Sieg:2005kd} of arbitrarily high order. Consequently, despite they are the simplest nontrivial operators one can compute in the theory, the authors expect that the results of this section can serve as a guiding principle when extending compact spin integrability results beyond the reach of Asymptotic Bethe Ansatz.

\acknowledgments
{We would like to thank Vladimir Kazakov, Christoph Sieg, Matteo Bertolini and Zohar Komargodski for the critical reading of the manuscript. We also want to thank Amit Sever, Zohar Komargodski, Ran Yacoby, Masazumi Honda, Lorenzo Di Pietro, Mikhail Isachenkov and Vladimir Narovlansky for useful discussions. 
This work was supported in part by Israel Science Foundation (grant number 1989/14), and by the ERC STG grant 335182.
The research of GT is supported by a Koshland Postdoctoral fellowship, partially financed by the Koshland Fundation.}

\appendix
\section{Finite order checks for the re-summed result}
\subsection{Perturbative expansion of $\gamma_{11}$}
\label{app:PerturbativePhi1}
\begin{figure}
	\begin{centering}
		\includegraphics[width=0.8\textwidth]{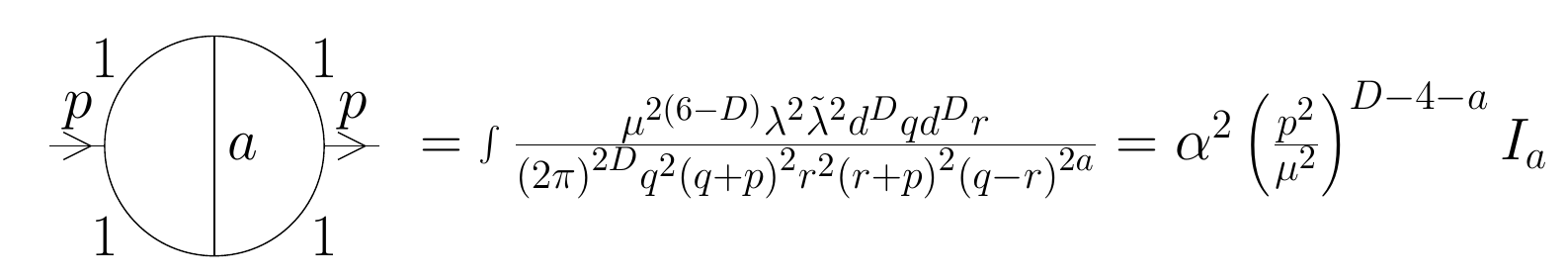}
		\par\end{centering}
	\caption{\label{fig:NomI}Elementary building block in the nested
		sequence depicted in figure \ref{fig:Diagram}, and the value
		of its amputated Feynman integral. When the additional $\left(p^{2}\right)^{-2}$ power coming from the external legs is taken into account,
			a generic diagram of the sequence becomes a product of $I_{2+2k\epsilon}$ factors in $D=6-2\epsilon$ dimensions, with the corresponding $\alpha$ and $p^2/\mu^2$ powers.}
\end{figure}

The perturbative expansion defined in figure \ref{fig:Diagram} could be written as:
\begin{equation}
\begin{array}{ccl}
\Sigma_{11}\left(p\right)& = &\mathcal{Z}_{11}^{-1} \left(1+\frac{\alpha^2\mu^{4\epsilon}}{p^{2\left(2\epsilon\right)}}I_{2}\left(1+\frac{\alpha^2\mu^{4\epsilon}}{p^{2\left(2\epsilon\right)}}I_{2+2\epsilon}\left(1+\frac{\alpha^2\mu^{4\epsilon}}{p^{2\left(2\epsilon\right)}}I_{2+4\epsilon}\left(1+\ldots\right)\right)\right)\right)=\\
& = &m^{2}\sum_{n=0}^{\infty}\alpha^{2n}\sum_{\ell=0}^{n}z_{n-\ell}\left(\frac{p^{2}}{\mu^{2}}\right)^{-2\,\epsilon\,\ell}\prod_{k=0}^{\ell-1}I_{2+2k\epsilon}
\end{array}
\label{eq:Expansion}\end{equation}
where $I_{a} \equiv I_{1,1,a}$ denotes the static factor in the
nested structure in the Feynman diagrams as figure \ref{fig:NomI}
illustrates, and we expanded $\mathcal{Z}_{11}$ according to the prescription of \ref{eq:ExpandZ}.

The exact expression for $I_{\alpha}$ in dimensional regularization
is simpler than the one for $I_{a,b,c}$, and was first derived in \cite{Kazakov:1984km}
using the uniqueness method\cite{Kazakov:1983pk}\footnote{See \cite{Grozin:2012xi} for an updated review on this type of diagrams and other ways of obtaining
	this result.}, which leads to
	
\begin{equation}
\begin{array}{c}
I_{a}=\frac{2\Gamma\left(2-\frac{D}{2}\right)\Gamma\left(\frac{D}{2}-1\right)^{2}\Gamma(a+3-D)\Gamma\left(\frac{D}{2}-a-1\right)}{\Gamma(a)}\left(\frac{\Delta}{\Gamma\left(3-\frac{D}{2}\right)\Gamma\left(\frac{3D}{2}-a-4\right)}-\frac{\Gamma(a)\cos\left(\pi\left(2-\frac{D}{2}\right)\right)}{\Gamma(D-2)}\right)\\
\Delta=\frac{\pFq{3}{2}{1,D-2,a+2-\frac{D}{2}}{3-\frac{D}{2},a+3-\frac{D}{2}}{-1}}{a+2-\frac{D}{2}}+\frac{\pFq{3}{2}{1,D-2,D-a-2}{3-\frac{D}{2},D-a-1}{-1}}{D-a-2}
\end{array}\label{eq:Exact}
\end{equation}
The recent publication \cite{Kotikov:2016rgs} presents new identities for hypergeometric functions that prove that \ref{eq:Exact} is indeed equivalent to the expression that \ref{eq:TheI} provides for $I_{1,1,a}$.

Having at our disposal the explicit form of all elements in \eqref{eq:Expansion},
we are now in position to series-expand them in $\epsilon$ in order
to fix the counterterms $\delta_{m^{2},n}$ and extract $\gamma_{\text{tr}\phi^{2}}$
from them. Schematically, this expansion proceeds, for $D=6-2\epsilon$,
in the following terms:
\begin{equation}
\begin{array}{c}
I_{2+2k\epsilon}=\sum_{j=0}^{\infty}c_{k,j}\epsilon^{j-1}\\
z_{n}=\sum_{k\le n}a_{n,k}\epsilon^{-k}\\
\left(\frac{p^{2}}{\mu^{2}}\right)^{-2\,\epsilon\,\ell}=\sum_{n}\frac{\left(-2\,\epsilon\,\ell\right)^{n}}{n!}\log^{n}\left(\frac{p^{2}}{\mu^{2}}\right)
\end{array}
\end{equation}
In the minimal subtraction scheme the $a_{n,k}$ coefficients are
iteratively identified from the set of $c_{n,k}$, and a nontrivial
sanity check on the computation is provided by the cancellation at
every order of divergent terms carrying logarithms of the momentum
by virtue of nondynamical counterterms fixed in previous orders. Let
us for illustrative purposes explicit here the first few terms of
this expansion:
\renewcommand{\arraystretch}{1.3}
	\begin{equation}
	\begin{array}{c}	
	\begin{array}{cc}
	c_{k,0}=\frac{1}{4\left(k+1\right)}; & c_{k,1}=\frac{13}{12}+\frac{7-4\gamma_{E}}{8+8k}-\zeta\left(3\right);
	\end{array}
	\\c_{k,2}=\frac{480 (-3 k+3 \gamma_E -7) (k+1) \zeta (3)-12 \pi ^4 (k+1)-60 \gamma_E  (26 k+47)+20 k (159 k+437)-30 \pi ^2+360 \gamma_E
   ^2+7135}{720 (k+1)}\\
   \begin{array}{ccc}
	a_{1,1}=-c_{0,0}=-\frac{1}{4}; & a_{2,2}=\frac{c_{0,0}^{2}}{2}=\frac{1}{32}; & a_{2,1}=\frac{c_{0,0}\left(c_{0,1}-2c_{1,1}\right)}{2}=\frac{1}{8}\left(\zeta\left(3\right)-\frac{13}{12}\right);
	\\ a_{3,3}=-\frac{c_{0,0}^3}{6}; & a_{3,2}=-\frac{3c_{0,0}\left(c_{0,1}-2c_{1,1}\right)}{6}; & a_{3,1}=-\frac{2c_{0,0}\left(c_{0,1}-2c_{1,1}\right)^2+c_{0,0}^2\left(c_{2,0}-4c_{2,1}+3c_{2,2}\right)}{6};
	\end{array}
	\end{array}
	\end{equation}
Additional coefficients can be generated computationally. The authors have used the Mathematica package HypExp\cite{Maitre:2005uu, Maitre:2007kp} to obtain analytic expressions for $c_{k,n>0}$.
\renewcommand{\arraystretch}{1.1}

The RG flow equations \ref{eq:RGFlow} imply that we can read $\gamma_{11}$ from the $\epsilon^{-1}$ divergence in $\mathcal{Z}_{11}^{-1}$, by adding an additional factor of $2n$ at order $\alpha^n$. The cancellation of higher order poles in $\epsilon$ has been verified numerically up to fifth order and provides a nontrivial check of the result for $I_a$. 
In addition, the perturbative expansion of the re-summed problem (obtained implicitly from equation \ref{eq:Transcendental} at $\alpha=0$):
\begin{equation}
\gamma_{11}= \sum_{n=1}^{\infty} 4n\, a_{n,1} \alpha^{2n}= 
-\alpha^{2}+\left(\zeta\left(3\right)-\frac{13}{12}\right)\alpha^4-
\left(\zeta(3)^2+\frac{8}{3}\zeta(3)+\frac{41}{18}\right)\alpha^6+
\ldots
\end{equation}

\subsection{Perturbative expansion of $\gamma_{12}$}
\label{app:PerturbativePhi1Phi2}
\begin{figure}
	\begin{centering}
		\includegraphics[width=0.8\textwidth]{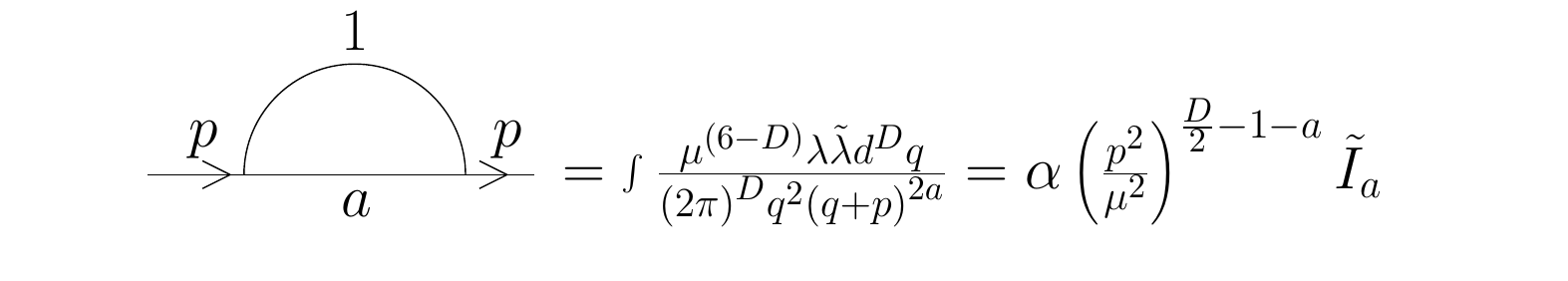}
		\par\end{centering}
	\caption{\label{fig:NomI2}Elementary building block in the nested
		sequence depicted in figure \ref{fig:Diagram2}, and value
		of its amputated Feynman integral. In this case the iterative factors of the sequence will be proportional to $I_{2+k\epsilon}$.}
\end{figure}

In the computation of the anomalous dimension of $\mbox{tr}\phi_1\phi_2$ the elementary piece of the nested structure (\ref{fig:NomI}) is replaced by the one represented in \ref{fig:NomI2}.
Notice the change in the powers of $\alpha$ and $p^2/\mu^2$, and the replacement of $I_{2+2k\epsilon}$ by $\tilde{I}_{2+k\epsilon}$, where the function $\tilde{I}$ is given by
\begin{equation}
\tilde{I}_a = 
\frac
{\Gamma \left(\frac{D}{2}-a\right)  \Gamma \left(\frac{D}{2}-1\right)  \Gamma \left(a+1-\frac{D}{2} \right)}
{\Gamma \left( a \right)            \Gamma \left(1 \right)             \Gamma \left(D-1-a \right)}
\end{equation}
In order to keep the notation as parallel as possible, we expand
\begin{equation}
\begin{array}{c}
\tilde{I}_{2+k\epsilon}=\sum_{j=0}^{\infty}\tilde{c}_{k,j}\epsilon^{j-1}
\end{array}
\end{equation}
When we insert this expansion into
\begin{equation}
\begin{array}{ccl}
\Sigma_{12}\left(p\right)& = &\mathcal{Z}_{12}^{-1}\left(1+\frac{\alpha\mu^{2\epsilon}}{p^{2\epsilon}}\tilde{I}_{2}\left(1+\frac{\alpha\mu^{2\epsilon}}{p^{2\epsilon}}\tilde{I}_{2+\epsilon}\left(1+\frac{\alpha\mu^{2\epsilon}}{p^{2\epsilon}}\tilde{I}_{2+2\epsilon}\left(1+\ldots\right)\right)\right)\right)=\\
& = &\sum_{n=0}^{\infty}\alpha^{n}\sum_{\ell=0}^{n}\tilde{z}_{n-\ell}\left(\frac{p^{2}}{\mu^{2}}\right)^{-\epsilon\,\ell}\prod_{k=0}^{\ell-1}\tilde{I}_{2+k\epsilon}
\end{array}
\label{eq:Expansion2}\end{equation}
we notice that the relations between $a$ and $c$ can be directly extrapolated to $\tilde{a}$ and $\tilde{c}$. The values of $\tilde{c}$ themselves, in contrast, have to be computed from the new $\tilde{I}$ function:
\begin{equation}
\begin{array}{ccc}
\tilde{c}_{k,0}=\frac{1}{2\left(k+1\right)}; & \tilde{c}_{k,1}= \frac{k-2 \gamma_E +4}{4 (k+1)}; & \tilde{c}_{k,2}=\frac{15 k^2-6 \gamma_E  k+42 k-\pi ^2+6 \gamma_E ^2-24 \gamma_E +48}{24 (k+1)}
\end{array}
\end{equation}
We obtain
\begin{equation}
\gamma_{12}=\sum_{n=1}^{\infty} 2n\, \tilde{a}_{n,1} \alpha^{n}=-\alpha -\frac{1}{4}\alpha^2-\frac{3}{8}\alpha^{3}-\ldots
\end{equation}
in concordance to the re-summed expression. The numerical verification has been performed up to 22 orders.
\interlinepenalty=10000 
\bibliographystyle{newutphys} 
\bibliography{GammaDef} 

\end{document}